\newcommand{\be}{\begin{equation}}
\newcommand{\ee}{\end{equation}}
\newcommand{\bea}{\begin{eqnarray}}
\newcommand{\eea}{\end{eqnarray}}
\newcommand{\nn}{\nonumber\\}
\newcommand{\p}[1]{(\ref{#1})}

\newcommand{\cJ}{{\cal J}}
\newcommand{\cT}{{\cal T}}
\newcommand{\cD}{{\cal D}}
\newcommand{\cW}{{\cal W}}
\documentclass[11pt]{article}
\begin{document}
\thispagestyle{empty}
\begin{flushright}
nlin.SI/0204063
\end{flushright}
\vspace{2cm}
\centerline{\large\bf Alternative dispersionless limit of N=2 supersymmetric
KdV-type hierarchies}
\vspace{1cm}
\begin{center}
Ashok Das $^{\ast}$, Sergey Krivonos $^{\dagger}$ and Ziemowit Popowicz
$^{\ddagger}$ \\
\vspace{1cm}
${}^{\ast}${\it Department of Physics and Astronomy, University of Rochester
\\ Rochester, NY 14627-0171, USA\\
$^{\dagger}$ Bogoliubov Laboratory of Theoretical Physics, JINR \\
141980, Dubna, Moscow region, Russia \\
$^{\ddagger}$ Institute of Theoretical Physics, University of Wroc\l aw\\
pl. M. Born 9, 50-204 Wroc\l aw, Poland}
\vspace{0.5cm}
\end{center}
\begin{center}
{\bf Abstract}
\end{center}

We present a systematic procedure for obtaining the dispersionless
limit of a class of $N=1$ supersymmetric systems starting from the Lax
description of their dispersive counterparts. This is achieved by
starting with an $N=2$ supersymmetric system and scaling the fields in
an alternative manner so as to maintain $N=1$ supersymmetry. We
illustrate our method by working out explicitly the examples of the
dispersionless supersymmetric two boson hierarchy and the
dispersionless supersymmetric Boussinesq hierarchy.
\vfill
\centerline{ April 2002}
\newpage

\section{Introduction} \vspace{-0.7cm}
Dispersionless integrable systems \cite{zakharov} have been studied,
in recent  years,
from various points of view. While the properties of such bosonic
models are quite well understood, much remains to be learnt about their
supersymmetric counterparts. For example, the Lax description of only
a handful of supersymmetric dispersionless systems have been
constructed, to date, by brute force \cite{barcelos,das} and there is
no  systematic
procedure for obtaining them starting from the corresponding
supersymmetric dispersive Lax
descriptions \cite{laberge,laxes,KS2,9}. Similarly, supersymmetric
dispersionless  systems often
have more conserved charges than their dispersive counterparts and we
do not yet know how to relate the new charges to the Lax function
itself.

In this letter, we take a modest step and show how one can obtain,
systematically, a Lax description for a select class of supersymmetric
dispersionless systems starting from their dispersive counterparts. We
describe in section {\bf 2} the basic procedure for taking the
dispersionless limit in the Lax description itself, within the context
of bosonic models. We work out some known examples to illustrate the
procedure and present the Lax description of some new bosonic models. In
section {\bf 3}, we extend  this method to
a class of supersymmetric models. We work out explicitly the example
of dispersionless supersymmetric two boson hierarchy starting from the
$N=2$ supersymmetric KdV hierarchy. In section {\bf 4}, we extend the
analysis and discuss the dispersionless supersymmetric Boussinesq
hierarchy and present a brief conclusion in section {\bf 5}.
\vspace{-0.7cm}
\section{Bosonic dispersionless systems}  \vspace{-0.7cm}
In this section, we describe the basic procedure of taking the
dispersionless limit of bosonic systems within the framework of the
Lax  description itself. Let us consider a general Lax operator of the
form
\begin{equation}
L = \partial^{n} + \sum_{m=1}^{\infty} A_{m} \partial^{n-m}
\end{equation}
where the coefficients, $A_{m}$'s, are functions of the dynamical variables of
the system which depend on the coordinates $(x,t)$. Let us assume that
the Lax equation
\begin{equation}
\frac{\partial L}{\partial t_k} = [(L^{k})_{\geq s}, L],\qquad s=0,1,2
\end{equation}
describes the dynamical system of equations. Here $()_{\geq s}$
represents the projection with respect to the powers of $\partial$.
In going to the
dispersionless limit, first of all, we replace $\partial\rightarrow
p$. Then, we scale $p\rightarrow \alpha p$ and all the basic dynamical
variables as $J_{i}\rightarrow (\alpha)^{i} J_{i}$ where $J_{i}$
represents the basic dynamical variables of the system with the
respective dimensions $i$. The Lax function which describes the
dispersionless system of equations is obtained to be \cite{zakharov}
\begin{equation}
{\cal L} = \lim_{\alpha\rightarrow\infty} \, \frac{1}{\alpha^n}\,
L_{\alpha}\label{limit}
\end{equation}
where $L_{\alpha}$ denotes the scaled Lax function. The dispersionless
equations are then obtained from the Lax equation
\begin{equation}
\frac{\partial {\cal L}}{\partial t_k} = - \left\{({\cal L}^{k})_{\geq
s}, {\cal L}\right\}\label{laxeq}
\end{equation}
where
\begin{equation}
\left\{A,B\right\} = \frac{\partial A}{\partial x}\frac{\partial
B}{\partial p} - \frac{\partial A}{\partial p} \frac{\partial
B}{\partial x}
\end{equation}
represents the Poisson bracket on the classical phase space.

Let us illustrate this procedure with a few examples. The reduction of the KdV
equation to its dispersionless limit is well known and, therefore, we
will not repeat it here. Rather, let us look at the two boson
hierarchy \cite{kupershmidt} described by the Lax operator
\begin{equation}
L = \partial - J + \partial^{-1} T
\end{equation}
with the nonstandard Lax equation given by
\begin{equation}
\frac{\partial L}{\partial t_{k}} = - \left[(L^{k})_{\geq 1}, L\right]
\end{equation}
Here $J$ and $T$ are dynamical field variables with  dimensions one
and two respectively. Therefore, under the scaling discussed
earlier, we have for the present case, $p\rightarrow \alpha p,
J\rightarrow \alpha J, T\rightarrow \alpha^{2}T$ and it follows that
in the dispersionless limit, the Lax operator goes into
\begin{equation}
{\cal L} = p - J + T p^{-1}
\end{equation}
and the Lax equation
\begin{equation}
\frac{\partial {\cal L}}{\partial t_{k}} = \left\{({\cal L}^{k})_{\geq
1}, {\cal L}\right\}
\end{equation}
describes the dispersionless system of equations.

The two boson hierarchy \cite{kupershmidt} can also be alternatively
described in terms
of the gauge equivalent Lax operator \cite{BX1}
\begin{equation}\label{boslax}
L=\partial - \frac{1}{\partial + J} \left(\frac{T}{2}\right)
\end{equation}
and the standard Lax equation
\begin{equation}\label{bosflow}
\frac{\partial L}{\partial t_k}= \left[ \left( L^k\right)_{\geq 0} , L\right]
\end{equation}
This description of the system of equations is more convenient from
the point of view of our subsequent discussions. We note that the second flow
\begin{equation}\label{bos2flow}
J_{t_2}=\left(  J_x + T- J^2 \right)_x\; , \quad T_{t_2}= - T_{xx}-2\left( JT\right)_x
\end{equation}
is the two boson equation which is also related to the nonlinear
Schr\"{o}dinger equation (NLS) \cite{BX1}. The third flow, on the
other hand, is obtained to be
\begin{equation}\label{bos3flow}
J_{t_3}= J_{xxx} +\left(J^3 -3JT-3JJ_x\right)_x \; , \quad T_{t_3} =T_{xxx} +
 3\left( J^2T-\frac{1}{2}T^2 +JT_x
   \right)_x
\end{equation}
and this coincides with the bosonic sector of the $N=2$ supersymmetric
KdV equation with $a=4$ \cite{laberge} after the transformations
\begin{equation}
J\rightarrow 2J\; ,\quad  T\rightarrow -2(T + J_{x})\; ,
\quad x\rightarrow i x \; ,\quad t\rightarrow  i t \; .
\end{equation}

In the present case, it is easy to check that, in the dispersionless
limit, the Lax function becomes
\begin{equation}\label{boslaxd}
{\cal L}=p - \frac{1}{p + J}\left(\frac{T}{2}\right) \;.
\end{equation}
The second and the third flow equations following from the Lax
equation
\begin{equation}
\frac{\partial {\cal L}}{\partial t_{k}} = - \left\{({\cal
L}^{k})_{\geq 0},{\cal L}\right\}
\end{equation}
are given by
\begin{eqnarray}
&& J_{t_2}=\left( T- J^2 \right)_x\; , \quad T_{t_2}= -2\left( JT\right)_x
 \label{bos2flowd} \\
&&J_{t_3}= \left(J^3 -3JT\right)_x \; , \quad T_{t_3} =3\left(
 J^2T-\frac{1}{2}T^2\right)_x \label{bos3flowd}
\end{eqnarray}
These are indeed the correct dispersionless limits of the two boson
hierarchy.

We would like to conclude this section by presenting a new system of
dispersionless equations. Let us consider a general Lax operator of
the form \cite{KS2}
\begin{equation}
L = \partial - \frac{1}{\partial^{m} + \sum_{i=1}^{m}
{J}_{i}\partial^{m-i}}\,\left(\sum_{i=1}^{m} \overline{J}_{i}
\partial^{m-i} \right)
\end{equation}
Here $J_{i},\overline{J}_{i}$ represent fields of dimension
$i$. It  can be checked that
\begin{equation}
\frac{\partial L}{\partial t_{k}} = \left[(L^{k})_{\geq 0},L\right]
\end{equation}
leads to a consistent set of dynamical equations. If we now follow the
earlier discussion, it is easy to check that, in the dispersionless
limit, the Lax function
\begin{equation}
{\cal L} = p - \frac{1}{p^{m} + \sum_{i=1}^{m} J_{i}
p^{m-i}}\,\left(\sum_{i=1}^{m} \overline{J}_{i}p^{m-i}\right)
\end{equation}
leads to the  system of dispersionless equations obtained from the Lax
equation
\begin{equation}\label{eq22}
\frac{\partial {\cal L}}{\partial t_{k}} = - \left\{({\cal
L}^{k})_{\geq 0}, {\cal L}\right\}
\end{equation}
\vspace{-0.7cm}
\section{Supersymmetric KdV and two boson hierarchy}  \vspace{-0.7cm}
In this section, we will generalize the ideas of the previous section
to construct the Lax description for a class of dispersionless
supersymmetric systems. In particular, we will work out in detail the
case of dispersionless $N=1$ supersymmetric two boson hierarchy starting
from the Lax description of the $N=2$ supersymmetric KdV hierarchy with
$a=4$ \cite{KS2} whose bosonic sector we have studied in the last section.
Let us
first note that the few dispersionless supersymmetric systems \cite{barcelos}
whose Lax descriptions have been constructed by brute force show that
although the
Lax function is defined in terms of superfields, it involves only bosonic
momenta  and that the conserved
charges are obtained from the bosonic residues of powers of the Lax
function. Furthermore, in the dispersionless limit, we know that
$\partial\rightarrow p$. However, the reduction of the fermionic covariant
derivative in the dispersionless limit is not well
understood. It is also already noted \cite{barcelos}
that a scaling of the fermionic covariant derivative is essential in order
to preserve supersymmetry in going to the dispersionless limit. In
view of the above mentioned difficulties, our strategy, as a first
step, is  to look at
supersymmetric systems which are described in terms of Lax operators
that involve only bosonic $\partial$ operators.

For one of the three known families of integrable supersymmetric hierarchies
with $N=2, W_n$ superalgebra as the second Hamiltonian structure,  the
Lax operators contain only bosonic operators of the forms \cite{KS2}:
\begin{equation}\label{superlax}
L_s=\partial -\left[ D {1\over\partial^s+\sum_{i=1}^{s}J_{i}\partial^{s-i}}
\overline{D}\left(\sum_{i=1}^{s}J_{i}\partial^{s-i}\right) \right] \;.
\end{equation}
Here, $s=0,1,2, \ldots$ and  $J_i$ are bosonic $N=2$ superfields of
dimensions $i$. Furthermore, the square brackets stand for
the fact  that
the $N=2$ supersymmetric fermionic covariant derivatives $D$ and
$\overline{D}$ defined to be
\begin{equation}
D = \frac{\partial}{\partial\theta} - \frac{\overline{\theta}}{2}
\partial \; ,\quad \overline{D} =
\frac{\partial}{\partial \overline{\theta}} - \frac{\theta}{2}
\partial
\end{equation}
act only on the superfields inside the brackets.

Let us consider the conventional dispersionless limit in the simplest case of
$N=2$ supersymmetric KdV hierarchy for which $s=1$.
The Lax operator, in this case, has the form
\begin{equation}\label{laxkdv}
L=\partial - \left[ D \frac{1}{\partial +J}\overline{D} J \right] \;.
\end{equation}
The second flow, following from this Lax operator, reads
\begin{equation}
J_{t_2} = \left( [D,\overline{D}]J   - J^{2}\right)_{x} \; .
\end{equation}
Under the rescaling $\partial_t\rightarrow \lambda\partial_t,
 \partial\rightarrow \lambda\partial,
(D,\overline{D})\rightarrow \lambda^{\frac{1}{2}} (D,\overline{D})$,
this equation will reduce to (as
$\lambda\rightarrow 0$)
\begin{equation}
J_{t_2} = - \left(J^{2}\right)_{x} \; .
\end{equation}
However,  this equation, despite being $N=2$ supersymmetric, is not very interesting.

A different possibility  is to rescale in a standard way
all the fields together with the fermionic derivatives
in the Lax operator \p{laxkdv}.
This leads to the following Lax function in the dispersionless limit:
\begin{equation}\label{laxkdvd0}
{\cal L} = p -\frac{\frac{1}{2}{\cal T}}{p+{\cal
 J}}-\frac{\frac{1}{2}\psi_1\psi_2}{(p+{\cal J})^2} \;,
\end{equation}
where we introduced the component fields:
\begin{equation}\label{comp}
{\cal J}=J|\;,\; \psi_1=(D+\overline{D})J|\;, \; \psi_2=(D-\overline{D})J|\; , \;
{\cal T} = \left[ D, \overline{D} \right] J| \;,
\end{equation}
and the restriction $|$ stands for keeping the $(\theta=\bar\theta=0)$ term.
One can check, that the second flow equations
\begin{eqnarray}\label{2flowd0}
&& {\cal J}_{t_2}=\left( {\cal T}- {\cal J}^2 \right)_x\; , \quad
{\cal T}_{t_2}= -2\left( {\cal J} {\cal T}-\psi_1\psi_2\right)_x\;,
\nonumber\\
&&  \left( \psi_1\right)_{t_2}=-2\left( {\cal J}\psi_1\right)_x \; , \quad
\left( \psi_2\right)_{t_2}=-2\left( {\cal J}\psi_2 \right)_x \;,
\end{eqnarray}
do not possess any supersymmetry at all.  They break even the $N=1$ supersymmetry. The same is also true
for higher flows.

Therefore  we propose the following alternative
approach to taking the dispersionless limit.
The main idea is to rescale the fermionic components of the  superfields $J_i$ differently
from the conventional method as
\begin{eqnarray}\label{susyscale}
&& \left. J_i\right| \rightarrow \alpha^i \left. J_i \right| \; , \;
\left.\left( D +\overline{D} \right)J_i\right|\rightarrow \alpha^i
    \left.\left( D +\overline{D} \right)J_i\right| \;, \\
\noalign{\vskip 4pt}%
&& \left.\left( D - \overline{D}\right)J_i\right| \rightarrow \alpha^{i+1}
   \left.\left( D - \overline{D} \right)J_i\right|\;, \left. \left[
D,\overline{D}\right]J_i\right| \rightarrow \alpha^{i+1} \left. \left[
D,\overline{D}\right]J_i\right| \;. \nonumber
\end{eqnarray}
It is clear that this unconventional, alternative rescaling
(\ref{susyscale})  will
explicitly  break the $N=2$  supersymmetry. However, a subset of
$N=1$ supersymmetry, generated by $(D+\overline{D})$ will survive and,
in the dispersionless limit, we will have a Lax description for an
$N=1$ supersymmetric system of equations.

Let us demonstrate in detail how all these  work for the Lax operator \p{laxkdv}.
According to our alternative procedure, the first
step  will consist of
representing  $\frac{1}{\partial +J}$ as
\begin{equation}
\frac{1}{\partial +J}\equiv \partial^{-1} +A_2\partial^{-2}+A_3\partial^{-3}+
  A_4\partial^{-4}+\ldots \;,
\end{equation}
where all the functions $A_n$ can be recursively calculated and the first
few have the explicit forms
\begin{equation}
A_2=-J\;,\; A_3=J^2+J_x\;, \;A_4=-J^3 -3JJ_x -J_{xx}\; ,\; \ldots\; .
\end{equation}
Thus, our Lax operator can also be written as
\begin{equation}\label{lax2}
L=\partial -\left[ D \left( \partial^{-1}\overline{D} J +A_2\partial^{-2}\overline{D} J+A_3\partial^{-3}\overline{D} J+
  A_4\partial^{-4}\overline{D} J +\ldots\right)\right] \;,
\end{equation}
and we should move the partial derivatives to the right in (\ref{lax2}).

The first non trivial term on the right hand side of (\ref{lax2})
generates an infinite series of terms when the derivative is moved to
the right, namely,
\begin{equation}\label{term1}
\partial^{-1} \left[ D\overline{D} J \right] \equiv \frac{1}{2}\left( {\cal T} -{\cal J}_x\right) \partial^{-1}-
\frac{1}{2}\left( {\cal T} -{\cal J}_x\right)_x \partial^{-2}+
\frac{1}{2}\left( {\cal T} -{\cal J}_x\right)_{xx} \partial^{-3}+\ldots \; .
\end{equation}
We may now replace $\partial\rightarrow p$ in the r.h.s. of (\ref{term1})
and  rescale
$$p\rightarrow \alpha p\;,\; {\cal J} \rightarrow \alpha {\cal J}\;,\;
\psi_1\rightarrow \alpha \psi_1 \; ,\;
\psi_2 \rightarrow \alpha^2 \psi_2 \; ,\;
{\cal T}\rightarrow \alpha^2 {\cal T} \;.$$
Then, it is easy to see that the only term from among  those in
(\ref{term1})  that will
contribute to $\lim_{\alpha\rightarrow\infty} \frac{1}{\alpha}
L_{\alpha}$ is
$$ \frac{1}{2}{\cal T} p^{-1} .$$
The second  term inside the square bracket in the right hand side of (\ref{lax2})
needs some more work:
\begin{eqnarray}\label{term21}
&&(DA_2)\partial^{-2}(\overline{D} J)+A_2\partial^{-2} (D\overline{D} J) \equiv
(DA_2)(\overline{D} J) \partial^{-2}-2(DA_2)(\overline{D} J)_x
\partial^{-3}+ \nonumber\\
&&\qquad\qquad A_2(D\overline{D} J)\partial^{-2}-
  2 A_2(D\overline{D} J)_x \partial^{-3}+\ldots \;,
\end{eqnarray}
where the dots stand for terms with $\partial^{-4}$ and higher. In the
scaling limit, only the following terms will survive
\begin{equation}\label{term22}
\frac{1}{2}\psi_1\psi_2p^{-2} -
\frac{1}{2}\psi_2\left(\psi_2\right)_x p^{-3}-
\frac{1}{2}{\cal J} {\cal T} p^{-2} \; .
\end{equation}
Continuing in a similar manner, we find the Lax operator in the
dispersionless  limit to be
\begin{equation}\label{laxkdvd}
{\cal L} = p -\frac{\frac{1}{2}{\cal T}}{p+{\cal
 J}}-\frac{\frac{1}{2}\psi_1\psi_2}{(p+{\cal J})^2}+
 \frac{\frac{1}{4}\psi_2(\psi_2)_x}{(p+{\cal J})^3}\;.
\end{equation}
The second flow, following from the Lax equation, has the form
\begin{eqnarray}\label{2flowd}
&& {\cal J}_{t_2}=\left( {\cal T}- {\cal J}^2 \right)_x\; , \quad
{\cal T}_{t_2}= -2\left( {\cal J} {\cal T}-\psi_1\psi_2\right)_x\;,
\nonumber\\
&&  \left( \psi_1\right)_{t_2}=\left( \left( \psi_2\right)_x-2{\cal
J}\psi_1\right)_x \; , \quad
\left( \psi_2\right)_{t_2}=-2\left( {\cal J}\psi_2 \right)_x \;.
\end{eqnarray}
These equations can also be easily rewritten in terms of $N=1$ superfields,
\begin{equation}
j={\cal J} +\theta\psi_1\; ,\quad \psi = \psi_2-\theta {\cal T}
\end{equation}
as
\begin{equation}
j_{t_2}=-\left( {\cal D}\psi+j^2\right)_x \; , \quad
\psi_{t_2}=-2\left(j\psi\right)_x\; ,
\end{equation}
where
$$ {\cal D}=\frac{\partial}{\partial\theta}-\theta\partial\;,\quad
{\cal D}^2=-\partial\;.$$

Let us note here that the Lax operator (\ref{laxkdvd}) is not new and
is gauge equivalent to the one which
has been constructed earlier by brute force in \cite{das}. However, we see
that it  can be systematically obtained from the
alternative dispersionless limit of the simplest of the Lax operators
in the family (\ref{superlax}).
\vspace{-0.7cm}
\section{Supersymmetric Boussinesq hierarchy}  \vspace{-0.7cm}
As a second example of our method, in this section, we will work out the
dispersionless limit starting from the $N=2$ supersymmetric Boussinesq
hierarchy with $\alpha = \frac{5}{2}$ \cite{9}, which is described
by the Lax operator (\ref{superlax}) with $s=2$ \cite{KS2}.
The Lax operator (\ref{superlax}), in this case, has the
explicit  form
\begin{equation}\label{blax}
L=\partial -\left[ D
\frac{1}{\partial^2+J_1\partial+J_2}\overline{D}\left(  J_1\partial +
J_2\right)\right]\;.
\end{equation}

Following our procedure, we will first rewrite
\begin{equation}\label{aux1}
\frac{1}{\partial^2+J_1\partial+J_2}=\partial^{-2}+A_1\partial^{-3}+A_2\partial^{-4}+\ldots \;,
\end{equation}
where all the  $A_n$'s can be easily calculated,
\begin{equation}\label{an}
A_1=-J_1\;,\;A_2=-J_2+2(J_1)_x+J_1^2\;,\; A_3=\left( 2J_2 -3(J_1)_x-\frac{5}{2}J_1^2\right)_x+2J_1J_2-J_1^3\;,\;\cdots
\end{equation}
With this, the first few terms in the square bracket in (\ref{blax})
have the form
\begin{eqnarray}\label{blax1}
\left[ D \frac{1}{\partial^2+J_1\partial+J_2}\overline{D}\left(
J_1\partial + J_2\right)\right] &=&
\left[ D \partial^{-2} \overline{D} \left( J_1\partial +J_2 \right) \right]+
  \left[ D A_1\partial^{-3} \overline{D} \left( J_1\partial +J_2
\right) \right]+ \\
&& \left[ D A_2 \partial^{-4} \overline{D} \left( J_1\partial +J_2
\right) \right]+
  \left[ D A_3\partial^{-5} \overline{D} \left( J_1\partial +J_2
\right) \right]+ \ldots \nonumber
\end{eqnarray}

Let us next introduce the components
\begin{eqnarray}\label{bcomp}
&&{\cal J}_1=J_1|\;,\; \psi_1=(D+\overline{D})J_1|\;,\;
\psi_2=(D-\overline{D} )J_1| \;, \; {\cal T}_1=\left[
D,\overline{D}\right]J_1| \;,\nonumber\\
&&{\cal T}_2=J_2|\;,\; \xi_1=(D+\overline{D})J_2|\;,\;
\xi_2=(D-\overline{D} )J_2| \;, \; {\cal W}=\left[
D,\overline{D}\right]J_2| \;,
\end{eqnarray}
which have the scaling behaviors:
\begin{equation}\label{bscale}
\left({\cal J}_1 , \psi_1 \right)\rightarrow \alpha \left({\cal J}_1 ,
\psi_1 \right)\;,\;
\left( \psi_2, {\cal T}_1, {\cal T}_2,\xi_1\right) \rightarrow
\alpha^2 \left( \psi_2, {\cal T}_1, {\cal T}_2,\xi_1\right)\;,\;
\left( \xi_2,{\cal W}\right)\rightarrow \alpha^3 \left( \xi_2,{\cal W}\right).
\end{equation}
We are now ready to find a Lax function  in the
dispersionless limit.

We can now have the fermionic  derivatives act on the fields in
(\ref{blax1}), move the partial derivatives to the right and  replace
$\partial\rightarrow p$. After this, it is easy to see that there will
be three types of terms that may survive in the limit (\ref{limit}):
\begin{eqnarray}\label{3terms}
{\cal L} &\equiv & p- {\cal A} -{\cal B} -{\cal C} \;, \nonumber\\
{\cal A} &\equiv & \frac{1}{2}\left( {\cal T}_1 p+{\cal
W}\right)\left(p^{-2}+A_1p^{-3}+A_2p^{-4}+\ldots \right) \\
{\cal B}& \equiv & \left(
(DA_1)p^{-3}+(DA_2)p^{-4}+\ldots\right)\left( (\overline{D} J_1)
p+(\overline{D} J_2)\right) \\
{\cal C}& \equiv &
\left(-3(DA_1)p^{-4}-4(DA_2)p^{-5}+\ldots\right)\left( (\overline{D}
J_1) p+(\overline{D} J_2)\right)_x \;.
\end{eqnarray}
Note that the expressions in the parenthesis  for ${\cal A},{\cal B}, {\cal C}$
contain terms with and without derivatives (see eq.\p{an}).
For terms of the types ${\cal A}$ and ${\cal C}$, there is no problem,
since in the dispersionless limit (scaling limit), only terms without
derivatives in $A_n$ (\ref{an}) contribute. In this case, we have:
\begin{equation}\label{acser}
{\cal A}= \frac{\frac{1}{2}\left({\cal T}_1 p+{\cal
W}\right)}{p^2+{\cal J}_1 p +{\cal T}_2} \; ,\quad
{\cal C}=-\left[D \frac{2p+J_1}{\left(p^2+J_1 p
+J_2\right)^2}\right]\left( (\overline{D} J_1)p+(\overline{D}
J_2)\right)_x\;.
\end{equation}
However, for terms of the type ${\cal B}$, the scaling require us to keep also the terms
linear in the first derivatives in all the $A_n$'s. This leads to
\begin{equation}\label{bser}
{\cal B}=\left[ D\left(\frac{1}{p^2+J_1 p +J_2}+ \frac{(2p+J_1)(J_1
p+J_2)_x}{\left(p^2+J_1 p +J_2\right)^3}\right)\right]
\left( (\overline{D} J_1) p +(\overline{D} J_2) \right) \;.
\end{equation}
In the dispersionless limit, the Lax function now becomes
\begin{eqnarray}\label{blaxd}
{\cal L}&=& p-\frac{\frac{1}{2}\left({\cal T}_1 p+{\cal
 W}\right)}{p^2+{\cal J}_1 p +{\cal T}_2}-
 \frac{\frac{1}{4}\psi_2\left(\psi_2 p+\xi_2\right)_x}{\left(p^2+{\cal
 J}_1 p +{\cal T}_2\right)^2}-
 \frac{\frac{1}{2}\left(\psi_1 p+\xi_1\right)\left(\psi_2
 p+\xi_2\right)}{\left(p^2+{\cal J}_1 p +{\cal T}_2\right)^2}+
 \nonumber\\
&& \frac{\frac{1}{4}(2p+{\cal J}_1)\left(\psi_2
 p+\xi_2\right)\left(\psi_2 p+\xi_2\right)_x}{\left(p^2+{\cal J}_1 p
 +{\cal T}_2\right)^3}+
\frac{\frac{1}{4}\left({\cal J}_1 p+{\cal T}_2\right)_x\psi_2
 \xi_2}{\left(p^2+{\cal J}_1 p +{\cal T}_2\right)^3} \;.
\end{eqnarray}
It is now easy to check that the Lax equation (\ref{eq22}) leads to the
dispersionless supersymmetric Boussinesq
hierarchy. Explicitly, the second flow of this hierarchy is given by
\begin{eqnarray}\label{beq}
&& \left( {\cal J}_1\right)_{t_2}=\left( 2{\cal T}_1+2{\cal T}_2-{\cal
J}_1^2\right)_x\;, \nonumber\\
&& \left( {\cal T}_1\right)_{t_2}=2\left( {\cal W} -{\cal J}_1{\cal
T}_1 +\psi_1\psi_2\right)_x \; , \;
 \left( {\cal T}_2\right)_{t_2}=-2({\cal J}_1)_x {\cal T}_2 +{\cal
J}_1({\cal T}_1)_x \;,\nonumber\\
&&\left( {\cal W}\right)_{t_2}=-2({\cal J}_1)_x{\cal W}-2({\cal
T}_1)_x{\cal T}_2+{\cal T}_1({\cal T}_1)_x+\psi_2(\psi_2)_{xx}+
    2\left( \xi_1(\psi_2)_x-\xi_2(\psi_1)_x\right) \;, \nonumber\\
&& \left( \psi_1\right)_{t_2}=2\left(\xi_1+(\psi_2)_x-{\cal
J}_1\psi_1\right)_x\; ,\;
\left( \psi_2\right)_{t_2}=2\left( \xi_2 -{\cal J}_1\psi_2\right)_x\;,
\nonumber\\
&& \left( \xi_1\right)_{t_2}=-2({\cal J}_1)_x\xi_1-2{\cal
T}_2(\psi_1)_x+({\cal T}_1)_x\psi_1+{\cal J}_1(\psi_2)_{xx}\;,
\nonumber\\
&&\left( \xi_2\right)_{t_2}=-2({\cal J}_1)\xi_2-2{\cal
T}_2(\psi_2)_x+({\cal T}_1)_x\psi_2\; .
\end{eqnarray}

This system of equations can also be rewritten in terms of $N=1$ superfields
\be\label{bn1sf}
j_1=\cJ_1+\theta\psi_1\;,\;\eta_1=\psi_2-\theta\cT_1\;,\; j_2=\cT_2+\theta\xi_1\; ,\;
  \eta_2=\xi_2-\theta\cW
\ee
as
\bea\label{bsf}
&& \left(j_1\right)_{t_2}=\left( -2\cD\eta_1+2j_2-j_1^2\right)_x\;\quad
\left( \eta_1\right)_{t_2}=2\left( \eta_2-j_1\eta_1\right)_x \;, \nn
&& \left( j_2\right)_{t_2}=-2(j_1)_x j_2 -j_1(\cD\eta_1)_x\; ,\;
\left(\eta_2\right)_{t_2}=-2(j_1)_x\eta_2-2j_2(\eta_1)_x-(\cD\eta_1 )_x \eta_1\;.
\eea
Thus, we explicitly demonstrate that our system possesses $N=1$ supersymmetry.
 \vspace{-0.7cm}
\section{Conclusion}  \vspace{-0.7cm}
In this paper, we have given a systematic derivation of the
dispersionless limit of a class of $N=1$ supersymmetric models starting from
the Lax description of their dispersive counterparts. This is achieved
by starting with the $N=2$ systems and making an alternative scaling
of the field variables which maintains only an $N=1$ supersymmetry.
This approach is motivated by the structure of the dispersionless limit of the pure
bosonic sectors of these $N=2$ systems which can not be extended to $N=2$
supersymmetry without introducing the additional bosonic fields.
We discuss our proposal explicitly within the context of the
supersymmetric two boson hierarchy where our starting point is the
$N=2$ supersymmetric KdV hierarchy. As a second example, we also work
out the dispersionless limit of the supersymmetric Boussinesq
hierarchy starting from the $N=2$ supersymmetric system.
\vspace{-0.7cm}
\section*{Acknowledgments}  \vspace{-0.7cm}
S.K. and Z.P. would like to thank the Department of Physics at the University of Rochester for
hospitality during the course of this work.
This work was supported in part by US DOE grant
no. DE-FG-02-91ER40685, NSF-INT-0089589, INTAS grant No. 00-00254, grant
DFG 436 RUS 113/669 as well as RFBR-CNRS grant No. 01-02-22005.

\end{document}